\documentclass[12pt]{iopart}
\usepackage{iopams}

\newcommand\ot{\mathop{\otimes}\limits}
\begin{document}

\title{Boundary correlation functions of the six-vertex model}

\author{N.M. Bogoliubov, A.G. Pronko and M.B. Zvonarev}

\address{
St. Petersburg Department of V.A. Steklov Mathematical Institute,
Fontanka 27, St.~Petersburg, 191011 Russia}

\begin{abstract}
We consider the six-vertex model on an $ N \times N $ square
lattice with the domain wall boundary conditions.
Boundary one-point correlation functions of the model are expressed
as determinants of $ N\times N $ matrices, generalizing the known result
for the partition function.
In the free fermion case the explicit answers are obtained.
The introduced correlation functions are closely related to the problem
of enumeration of alternating sign matrices and domino tilings.
\end{abstract}




\section{Introduction}

The six-vertex model was studied for both periodic
\cite{B-82,LW-72} and fixed boundary conditions
\cite{BO-89,BBRY-95,E-99,BO-98}.
The particular example of fixed boundary conditions \cite{B-87} of
the model on an $N\times N$ square lattice is the so-called domain wall
boundary conditions (DWBC) \cite{K-82}.
Under special restrictions on the vertex weights this model is related to the
enumeration of alternating sign matrices \cite{EKLP-92,Ku-96}
and domino tilings of Aztec diamonds \cite{EKLP-92}.

The model with DWBC originally appeared in the
context of investigation of norms of the Bethe states in the framework
of the Quantum Inverse Scattering Method (QISM) \cite{KBI-93}.
In the last decade the six-vertex
model with DWBC has found interesting applications in different
fields of physics and mathematics \cite{J-00,KJ-00,IKK-00,NikS-00,L-99}
due to the results of the papers
\cite{I-87,ICK-92}, where the determinant formula for the partition function
has been obtained and proved. This determinant formula 
allowed to solve several problems in combinatorics \cite{Ku-96}
which were standing for a long time \cite{Br-99}.

A wide range of problems such as refined enumeration of
alternating sign matrices (ASM)
\cite{Z-96} and the arctic circle theorem \cite{CEP-96,JPS-98} can be
solved only if the correlation functions of the model are known.
In general, the calculation of the correlation functions
is a more complicated problem than that of the partition function.
Additional difficulties may arise due to the lack of translation invariance
caused by the fixed boundary conditions.

In this paper we will consider two kinds of one-point boundary
correlation functions of the six-vertex model with DWBC.
The function of the first kind, $G_N^{(M)}$,
is the local state probability on the boundary vertical edge
and it may be called ``boundary spontaneous polarization''.
The function of the second kind, $H_N^{(M)}$,
describes the probability of the vertex being in the specific state.
For the model on an $N\times N$ square lattice we obtain
representations for these correlation functions as determinants of
$N\times N$ matrices.
These correlation functions are the generalization of a boundary
correlation function considered in \cite{BKZ-02}.

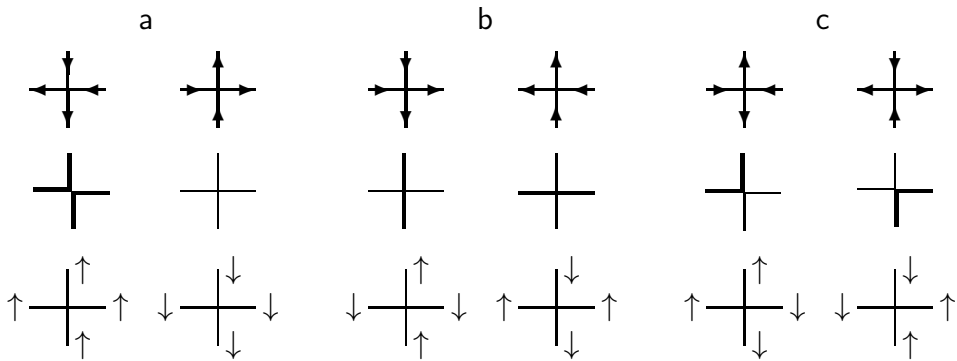
\begin{figure}[b]
\unitlength=1mm
\begin{center}
\begin{picture}(124,15)
\thicklines
\put(7,0){\line(0,1){10}}
\put(7.05,10){\vector(0,-1){3}}
\put(7.05,5){\vector(0,-1){5}}
\put(2,5){\line(1,0){10}}
\put(7,5){\vector(-1,0){5}}
\put(12,5){\vector(-1,0){3}}
\put(16.5,13){$ {\sf a} $}
\put(27,0){\line(0,1){10}}
\put(27.05,5){\vector(0,1){5}}
\put(27.05,0){\vector(0,1){3}}
\put(22,5){\line(1,0){10}}
\put(22,5){\vector(1,0){3}}
\put(27,5){\vector(1,0){5}}
\put(52,0){\line(0,1){10}}
\put(52,10){\vector(0,-1){3}}
\put(52,5){\vector(0,-1){5}}
\put(47,5){\line(1,0){10}}
\put(47,5){\vector(1,0){3}}
\put(52,5){\vector(1,0){5}}
\put(61.5,13){$ {\sf b } $}
\put(72,0){\line(0,1){10}}
\put(72,5){\vector(0,1){5}}
\put(72,0){\vector(0,1){3}}
\put(67,5){\line(1,0){10}}
\put(72,5){\vector(-1,0){5}}
\put(77,5){\vector(-1,0){3}}
\put(97,0){\line(0,1){10}}
\put(97,5){\vector(0,1){5}}
\put(97,5){\vector(0,-1){5}}
\put(92,5){\line(1,0){10}}
\put(92,5){\vector(1,0){3}}
\put(102,5){\vector(-1,0){3}}
\put(106.5,13){$ {\sf c} $}
\put(117,0){\line(0,1){10}}
\put(117,10){\vector(0,-1){3}}
\put(117,0){\vector(0,1){3}}
\put(112,5){\line(1,0){10}}
\put(117,5){\vector(-1,0){5}}
\put(117,5){\vector(1,0){5}}
\end{picture}
\end{center}
\begin{center}
\begin{picture}(124,10)
\linethickness{0.5mm}
\put(7.25,5.25){\line(0,1){4.75}}
\put(7.5,5.25){\line(-1,0){5}}
\put(7.75,4.75){\line(0,-1){4.75}}
\put(7.5,4.75){\line(1,0){5}}
\thinlines
\put(27,0){\line(0,1){10}}
\put(22,5){\line(1,0){10}}
\linethickness{0.5mm}
\put(51.75,0){\line(0,1){10}}
\thinlines
\put(47,5){\line(1,0){10}}
\thinlines
\put(72,0){\line(0,1){10}}
\linethickness{0.5mm}
\put(67,4.75){\line(1,0){10}}
\linethickness{0.5mm}
\put(96.75,5){\line(0,1){5}}
\put(96.9,5){\line(-1,0){5}}
\thinlines
\put(97,4.75){\line(0,-1){5}}
\put(97,4.7){\line(1,0){4.7}}
\put(117,5.25){\line(0,1){4.75}}
\put(117,5.25){\line(-1,0){5}}
\linethickness{0.5mm}
\put(117.25,5){\line(0,-1){4.75}}
\put(117,5){\line(1,0){5}}
\end{picture}
\end{center}
\begin{center}
\begin{picture}(124,14)
\thinlines
\put(7,2){\line(0,1){10}}
\put(8,11){$\uparrow$}
\put(8,1){$\uparrow$}
\put(2,7){\line(1,0){10}}
\put(-1,6){$\uparrow$}
\put(13,6){$\uparrow$}
\put(27,2){\line(0,1){10}}
\put(28,11){$\downarrow$}
\put(28,1){$\downarrow$}
\put(22,7){\line(1,0){10}}
\put(19,6){$\downarrow$}
\put(33,6){$\downarrow$}
\put(52,2){\line(0,1){10}}
\put(53,11){$\uparrow$}
\put(53,1){$\uparrow$}
\put(47,7){\line(1,0){10}}
\put(44,6){$\downarrow$}
\put(58,6){$\downarrow$}
\put(72,2){\line(0,1){10}}
\put(73,11){$\downarrow$}
\put(73,1){$\downarrow$}
\put(67,7){\line(1,0){10}}
\put(64,6){$\uparrow$}
\put(78,6){$\uparrow$}
\put(97,2){\line(0,1){10}}
\put(98,11){$\uparrow$}
\put(98,1){$\downarrow$}
\put(92,7){\line(1,0){10}}
\put(89,6){$\uparrow$}
\put(103,6){$\downarrow$}
\put(117,2){\line(0,1){10}}
\put(118,11){$\downarrow$}
\put(118,1){$\uparrow$}
\put(112,7){\line(1,0){10}}
\put(109,6){$\downarrow$}
\put(123,6){$\uparrow$}
\end{picture}
\end{center}
\caption{The six allowed types of vertices: in terms of arrows (first row),
in terms of lines (second row) and in terms of spins (third row).}
\label{vertices}
\end{figure}

There are three
convenient ways for description of the six-vertex model:
(i) in terms of arrows pointing into and away from each vertex;
(ii) in terms of lines flowing through the vertices;
(iii) in terms of spins on the edges.
The six types of vertices allowed in the model are plotted in
Figure~\ref{vertices}.
A statistical (vertex) weight corresponds to each type of vertex.
We consider the six-vertex
model with the vertex weights being invariant under the simultaneous
reversal of all arrows. Hence, there are three
different vertex weights, ${\sf a}$, ${\sf b}$ and ${\sf c}$, see Figure 1.

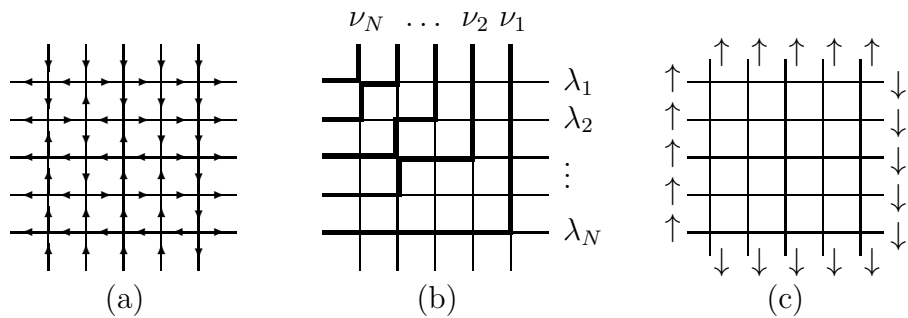
\begin{figure}[ht]
\unitlength=1mm
\begin{center}
\begin{picture}(35,35)
\put(0,5){\line(1,0){30}}
\put(5,5){\vector(-1,0){3.5}}
\put(10,5){\vector(-1,0){3.5}}
\put(15,5){\vector(-1,0){3.5}}
\put(20,5){\vector(-1,0){3.5}}
\put(25,5){\vector(-1,0){3.5}}
\put(25,5){\vector(1,0){3.5}}
\put(0,10){\line(1,0){30}}
\put(5,10){\vector(-1,0){3.5}}
\put(10,10){\vector(-1,0){3.5}}
\put(10,10){\vector(1,0){3.5}}
\put(15,10){\vector(1,0){3.5}}
\put(20,10){\vector(1,0){3.5}}
\put(25,10){\vector(1,0){3.5}}
\put(0,15){\line(1,0){30}}
\put(5,15){\vector(-1,0){3.5}}
\put(10,15){\vector(-1,0){3.5}}
\put(15,15){\vector(-1,0){3.5}}
\put(20,15){\vector(-1,0){3.5}}
\put(20,15){\vector(1,0){3.5}}
\put(25,15){\vector(1,0){3.5}}
\put(0,20){\line(1,0){30}}
\put(5,20){\vector(-1,0){3.5}}
\put(5,20){\vector(1,0){3.5}}
\put(15,20){\vector(-1,0){3.5}}
\put(15,20){\vector(1,0){3.5}}
\put(20,20){\vector(1,0){3.5}}
\put(25,20){\vector(1,0){3.5}}
\put(0,25){\line(1,0){30}}
\put(5,25){\vector(-1,0){3.5}}
\put(10,25){\vector(-1,0){3.5}}
\put(10,25){\vector(1,0){3.5}}
\put(15,25){\vector(1,0){3.5}}
\put(20,25){\vector(1,0){3.5}}
\put(25,25){\vector(1,0){3.5}}
\put(5,0){\line(0,1){30}}
\put(5.05,0){\vector(0,1){3.5}}
\put(5.05,5){\vector(0,1){3.5}}
\put(5.05,10){\vector(0,1){3.5}}
\put(5.05,15){\vector(0,1){3.5}}
\put(5.05,25){\vector(0,-1){3.5}}
\put(5.05,30){\vector(0,-1){3.5}}
\put(10,0){\line(0,1){30}}
\put(10.05,0){\vector(0,1){3.5}}
\put(10.05,5){\vector(0,1){3.5}}
\put(10.05,15){\vector(0,-1){3.5}}
\put(10.05,20){\vector(0,-1){3.5}}
\put(10.05,20){\vector(0,1){3.5}}
\put(10.05,30){\vector(0,-1){3.5}}
\put(15,0){\line(0,1){30}}
\put(15.05,0){\vector(0,1){3.5}}
\put(15.05,5){\vector(0,1){3.5}}
\put(15.05,10){\vector(0,1){3.5}}
\put(15.05,15){\vector(0,1){3.5}}
\put(15.05,25){\vector(0,-1){3.5}}
\put(15.05,30){\vector(0,-1){3.5}}
\put(20,0){\line(0,1){30}}
\put(20.05,0){\vector(0,1){3.5}}
\put(20.05,5){\vector(0,1){3.5}}
\put(20.05,10){\vector(0,1){3.5}}
\put(20.05,20){\vector(0,-1){3.5}}
\put(20.05,25){\vector(0,-1){3.5}}
\put(20.05,30){\vector(0,-1){3.5}}
\put(25,0){\line(0,1){30}}
\put(25.05,5){\vector(0,-1){3.5}}
\put(25.05,10){\vector(0,-1){3.5}}
\put(25.05,15){\vector(0,-1){3.5}}
\put(25.05,20){\vector(0,-1){3.5}}
\put(25.05,25){\vector(0,-1){3.5}}
\put(25.05,30){\vector(0,-1){3.5}}
\put(12.5,-5){(a)}
\end{picture}
\hskip .2in
\begin{picture}(35,35)
\put(0,5){\line(1,0){30}}
\put(0,10){\line(1,0){30}}
\put(0,15){\line(1,0){30}}
\put(0,20){\line(1,0){30}}
\put(0,25){\line(1,0){30}}
\put(5,0){\line(0,1){30}}
\put(10,0){\line(0,1){30}}
\put(15,0){\line(0,1){30}}
\put(20,0){\line(0,1){30}}
\put(25,0){\line(0,1){30}}
\linethickness{0.5mm}
\put(0,5){\line(1,0){25.25}}
\put(25,30){\line(0,-1){25.25}}
\put(0,10){\line(1,0){10.25}}
\put(20,30){\line(0,-1){15.25}}
\put(20.25,14.75){\line(-1,0){10}}
\put(10.25,9.75){\line(0,1){5.25}}
\put(0,15.25){\line(1,0){10}}
\put(15,30){\line(0,-1){10.25}}
\put(15,20){\line(-1,0){5.5}}
\put(9.75,20){\line(0,-1){5}}
\put(0,20){\line(1,0){5.25}}
\put(10,30){\line(0,-1){5.25}}
\put(10.25,24.75){\line(-1,0){5}}
\put(5.25,25){\line(0,-1){5.25}}
\put(0,25.25){\line(1,0){5}}
\put(4.75,30){\line(0,-1){5}}
\put(3.5,32.5){$\nu_N$}
\put(11,32.5){$\dots$}
\put(18.5,32.5){$\nu_2$}
\put(23.5,32.5){$\nu_1$}
\put(32,4){$\lambda_N$}
\put(32,11){$\vdots$}
\put(32,19){$\lambda_2$}
\put(32,24){$\lambda_1$}
\put(12.5,-5){(b)}
\end{picture}
\hskip .4in
\begin{picture}(35,35)
\put(2,5){\line(1,0){26}}
\put(2,10){\line(1,0){26}}
\put(2,15){\line(1,0){26}}
\put(2,20){\line(1,0){26}}
\put(2,25){\line(1,0){26}}
\put(5,2){\line(0,1){26}}
\put(10,2){\line(0,1){26}}
\put(15,2){\line(0,1){26}}
\put(20,2){\line(0,1){26}}
\put(25,2){\line(0,1){26}}
\put(5.5,28){$\uparrow$}
\put(10.5,28){$\uparrow$}
\put(15.5,28){$\uparrow$}
\put(20.5,28){$\uparrow$}
\put(25.5,28){$\uparrow$}
\put(29,3.5){$\downarrow$}
\put(29,8.5){$\downarrow$}
\put(29,13.5){$\downarrow$}
\put(29,18.5){$\downarrow$}
\put(29,23.5){$\downarrow$}
\put(5.5,0){$\downarrow$}
\put(10.5,0){$\downarrow$}
\put(15.5,0){$\downarrow$}
\put(20.5,0){$\downarrow$}
\put(25.5,0){$\downarrow$}
\put(-1,4.5){$\uparrow$}
\put(-1,9.5){$\uparrow$}
\put(-1,14.5){$\uparrow$}
\put(-1,19.5){$\uparrow$}
\put(-1,24.5){$\uparrow$}
\put(12.5,-5){(c)}
\end{picture}
\end{center}
\caption{One of the possible configurations in the model with DWBC:
(a) in terms of arrows; (b) in terms of lines.
(c) DWBC in terms of spins.}
\label{DWBC}
\end{figure}

The domain wall boundary conditions imply that all arrows on the top
and bottom of the lattice are pointing inward while
all arrows on the left and right boundaries are pointing outward,
see Figure~\ref{DWBC}(a).
It means that the solid lines flow from the top
of the lattice to the left boundary, see Figure~\ref{DWBC}(b).
In Figures~\ref{DWBC}(a) and \ref{DWBC}(b) one of the
possible arragements is presented as well.
The domain wall boundary conditions in terms of spins are shown
in Figure~\ref{DWBC}(c).

In the inhomogeneous model the wertex weights
${\sf a}$, ${\sf b}$ and ${\sf c}$ are site dependent.
To introduce this dependence we will use two sets of the variables
$\{\lambda_\alpha\}$ and $\{\nu_k\}$ that are in one to one
correspondence with the set of lines.
The rows will be enumerated by Greek indices $ \alpha=1,\dots,N $ and
the variable $\lambda_\alpha$ corresponds to $ \alpha$-th row;
the columns will be enumerated by Latin indices $ k=1,\dots,N $ and
the variable $ \nu_k $ corresponds to $ k $-th column.
This correspondence is shown in Figure~\ref{DWBC}(b).
Each statistical weight associated with the
vertex lying at the intersection of $\alpha$-th row and $k$-th column
will depend on the pair of variables $(\lambda_\alpha$, $\nu_k)$. The
parametrization that allows one to apply QISM is
\begin{equation} \label{abc}
\eqalign{
{\sf a}(\lambda_\alpha,\nu_k)=\sinh(\lambda_\alpha-\nu_k+\eta),\\
{\sf b}(\lambda_\alpha,\nu_k)=\sinh(\lambda_\alpha-\nu_k-\eta),\\
{\sf c}(\lambda_\alpha,\nu_k)=\sinh 2\eta.
}
\end{equation}
In the homogeneous limit all $ \lambda_\alpha \to \lambda $, and
all $ \nu_k \to \nu $.
All positive values of the vertex weights,
up to an overall scaling transformation,
may be obtained by choosing both $ \lambda - \nu $ and $ \eta $
either real or pure imaginary.

Our calculations are based on the Quantum Inverse Scattering Method (QISM),
reviewed briefly in Section 2. In Section 3
we derive the reduction formulae for the boundary correlation functions
$ G_{N}^{(M)} $ and $ H_{N}^{(M)} $. The recursion relation for the partition
function follows from these formulae as a particular case.
The determinant representation for $ G_{N}^{(M)} $ and $ H_{N}^{(M)} $
is obtained from these reduction formulae in Section 4. In the
free fermion case the homogeneous limit for these boundary correlation
functions is calculated explicitly in Section 5, while the general case
of the homogeneous limit is considered in Section 6.

\section{Formulation of the model within QISM formalism}

To apply the Quantum Inverse Scattering Method \cite{KBI-93}
we use the spin description of the model.
With each vertical line (column) and horizontal line (row)
one associates the space $\mathbb{C}^2$,
with spin up and spin down states forming a natural basis in this space.
The total space of the vertical lines is
$ {\cal V}=(\mathbb{C}^2)^{\otimes N} $
and the total space of the horizontal lines is
$ {\cal H}=(\mathbb{C}^2)^{\otimes N} $.
With each vertex of the lattice one associates an operator
acting in the full space ${\cal V}\otimes {\cal H}$.
This operator is called $\sf L$-operator
and it acts nontrivially only in a single
horizontal space $\mathbb{C}^2$ and in a single vertical space
$\mathbb{C}^2$,
while in all other spaces it acts as the unity operator.
To distinguish the spaces in which the $\sf L$-operator acts nontrivially
one can label it as ${\sf L}_{\alpha k}$ and associate it with the vertex
being the intersection of $ \alpha $-th row and $ k $-th column. The
matrix elements of the ${\sf L}$-operator
(which is $2^{2N}\times 2^{2N}$ matrix) are either zeros or
functions ${\sf a}(\lambda_\alpha,\nu_k)$,
${\sf b}(\lambda_\alpha,\nu_k)$, ${\sf c}(\lambda_\alpha,\nu_k)$,
defined in (\ref{abc}).
Hence, the ${\sf L}$-operator ${\sf L}_{\alpha k}$ is the
function of $\lambda_\alpha$ and $\nu_k$,
${\sf L}_{\alpha k}(\lambda_\alpha,\nu_k)$. Since the ${\sf L}$-operator
acts nontrivially only in the direct product of a pair of two-dimensional
spaces, all its elements may be written in the compact form
as
\begin{equation} 
{\sf L}_{\alpha k} (\lambda_\alpha,\nu_k)
=\pmatrix{{\sf a}\,(\lambda_\alpha,\nu_{k}) & 0 & 0 & 0 \cr
0 & {\sf b}\,(\lambda_\alpha,\nu_{k})
& {\sf c}\,(\lambda_\alpha,\nu_{k}) & 0 \cr
0 & {\sf c}\,(\lambda_\alpha,\nu_{k})
& {\sf b}\,(\lambda_\alpha,\nu_{k}) & 0 \cr
0 & 0 & 0 & {\sf a}\,(\lambda_\alpha,\nu_{k})}_{[\alpha k]}.
\end{equation}
This is the matrix with respect to $\alpha$-th copy
of $\mathbb{C}^2$ in $\cal H$ and $k$-th copy in $\cal V$, with the
matrix elements being trivial matrices in the rest copies of
$\mathbb{C}^2$ in $\cal V$ and in $\cal H$.
One can write the $\sf L$-operator
in the alternative form with the separated ``horizontal'' and ``vertical''
spaces, namely, as a matrix with respect to
$\alpha$-th copy of $\mathbb{C}^2$ in $\cal H$ with the operator
matrix elements acting nontrivially only in $k$-th copy
of $\mathbb{C}^2$ in $\cal V$
\begin{equation} \label{L}
{\sf L}_{\alpha k} (\lambda_\alpha,\nu_k)
=\pmatrix{
\sinh(\lambda_\alpha-\nu_k+\eta\sigma_k^{z})
& \sigma_k^{-}\sinh 2\eta \cr
\sigma_k^{+}\sinh 2\eta
& \sinh(\lambda_\alpha-\nu_k-\eta\sigma_k^{z})
}_{[\alpha]},
\end{equation}
where $\sigma_k^{z}$, $\sigma_k^{\pm}=\case 12
(\sigma_k^x\pm\rmi\sigma_k^y)$ are Pauli matrices.

The main object of QISM is the ``vertical''
monodromy matrix ${\sf T}_\alpha(\lambda_\alpha)$ which
is defined as the ordered matrix product of the $\sf L$-operators
along $\alpha$-th horizontal line
\begin{equation} \label{T}
{\sf T}_\alpha(\lambda_\alpha) =
{\sf L}_{\alpha N} (\lambda_\alpha,\nu_N)\dots
{\sf L}_{\alpha 1} (\lambda_\alpha,\nu_1) =
\pmatrix{A(\lambda_\alpha) & B(\lambda_\alpha) \cr
C(\lambda_\alpha) & D(\lambda_\alpha)}_{[\alpha]}.
\end{equation}
All entries of the monodromy matrix ${\sf T}_\alpha(\lambda_\alpha)$
are operators acting in $\cal V$ and they depend on the variables
$\nu_1,\dots,\nu_N$, i.e.,
$ A(\lambda) = A(\lambda; \{\nu_k\}_{k=1}^N) $, etc.
Obviously, instead of the ``vertical'' monodromy matrix one may use the
``horizontal'' one, which is the ordered product of the $\sf L$-operators
along the vertical line.

The Quantum Inverse Scattering Method is based on
the intertwining relation for the $\sf L$-operators:
\begin{equation} \label{RLL}
\fl
{\sf R}_{\alpha\beta}(\lambda_\alpha,\lambda_\beta)\,
{\sf L}_{\alpha k}(\lambda_\alpha,\nu_k)\,
{\sf L}_{\beta k}(\lambda_\beta,\nu_k)
= {\sf L}_{\beta k}(\lambda_\beta,\nu_k)\,
{\sf L}_{\alpha k}(\lambda_\alpha,\nu_k)\,
{\sf R}_{\alpha\beta}(\lambda_\alpha,\lambda_\beta),
\qquad
\alpha\ne\beta.
\end{equation}
The ${\sf R}$-matrix ${\sf R}_{\alpha\beta}(\lambda,\lambda')$
acts nontrivially in the direct product of $\alpha$-th and $\beta$-th
horizontal spaces and are given by
\begin{equation} 
${\sf R}$_{\alpha\beta}(\lambda,\lambda')
=\pmatrix{f(\lambda',\lambda) & 0 & 0 & 0 \cr
0 & 1 & g(\lambda',\lambda) & 0 \cr
0 & g(\lambda',\lambda)& 1 & 0 \cr
0 & 0 & 0 & f(\lambda',\lambda)}_{[\alpha\beta]},
\end{equation}
where the functions $f(\lambda',\lambda)$ and $g(\lambda',\lambda)$ are
\begin{equation} \label{fg}
f(\lambda',\lambda) =
\frac{\sinh(\lambda-\lambda'+2\eta)}{\sinh(\lambda-\lambda')}, \qquad
g(\lambda',\lambda) = \frac{\sinh 2\eta}{\sinh(\lambda-\lambda')}.
\end{equation}
This is the so-called trigonometric $\sf R$-matrix \cite{KBI-93},
which satisfies the Yang-Baxter equation
\begin{eqnarray} 
\fl
{\sf R}_{\alpha\beta}(\lambda_\alpha,\lambda_\beta)\,
{\sf R}_{\alpha \gamma}(\lambda_\alpha,\lambda_\gamma)\,
{\sf R}_{\beta \gamma}(\lambda_\beta,\lambda_\gamma)
\nonumber\\
\lo
= {\sf R}_{\beta \gamma}(\lambda_\beta,\lambda_\gamma)\,
{\sf R}_{\alpha \gamma}(\lambda_\alpha,\lambda_\gamma)\,
{\sf R}_{\alpha\beta}(\lambda_\alpha,\lambda_\beta),
\qquad
\alpha\ne\beta\ne\gamma.
\end{eqnarray}
Due to relation (\ref{RLL}) and commutativity of the matrix elements
of $\sf L$-operator (\ref{L}) at different lattice sites
one has the intertwining relation for the monodromy matrix:
\begin{equation} \label{RTT}
{\sf R}_{\alpha\beta}(\lambda_\alpha,\lambda_\beta)\,
{\sf T}_{\alpha}(\lambda_\alpha)\,
{\sf T}_{\beta}(\lambda_\beta)
= {\sf T}_{\beta}(\lambda_\beta)\,
{\sf T}_{\alpha}(\lambda_\alpha)\,
{\sf R}_{\alpha\beta}(\lambda_\alpha,\lambda_\beta),
\qquad
\alpha\ne\beta.
\end{equation}
Equation (\ref{RTT}) defines the commutation relations for the
operators entering the monodromy matrix. The complete list
of these relations can be found, e.g., in \cite{KBI-93}.
For our purposes we need only two of them:
\begin{eqnarray} \label{commAB}
\eqalign{
A(\lambda)\, B(\lambda') =
f(\lambda,\lambda')\, B(\lambda')\, A(\lambda) +
g(\lambda',\lambda)\, B(\lambda)\, A(\lambda'),
\\
B(\lambda)\, B(\lambda') = B(\lambda')\, B(\lambda).
}
\end{eqnarray}
As generating vector in the space ${\cal V}$ it is convenient to use
the state either with all spins up or with all spins down
\begin{equation} \label{ud}
\fl
|\Uparrow\,\rangle = \otimes_{k=1}^{N} |\uparrow\,\rangle_{k}
= \otimes_{k=1}^{N} \pmatrix{1 \cr 0}_{[k]},\qquad
|\Downarrow\,\rangle = \otimes_{k=1}^{N} |\downarrow\,\rangle_{k}
= \otimes_{k=1}^{N} \pmatrix{0 \cr 1}_{[k]}.
\end{equation}
These vectors are annihilated by the operators $C(\lambda)$ and
$B(\lambda)$, respectively,
\begin{equation} \label{die}
C(\lambda)\,|\Uparrow\,\rangle = 0,\qquad
B(\lambda)\,|\Downarrow\,\rangle = 0,
\end{equation}
and they are eigenvectors of the operators $A(\lambda)$ and
$D(\lambda)$
\begin{eqnarray} \label{updown}
\eqalign{
A(\lambda)\,|\Uparrow\,\rangle =
a(\lambda)\,|\Uparrow\,\rangle, \qquad
D(\lambda)\,|\Uparrow\,\rangle =
d(\lambda)\,|\Uparrow\,\rangle,
\\
A(\lambda)\,|\Downarrow\,\rangle =
d(\lambda)\,|\Downarrow\,\rangle, \qquad
D(\lambda)\,|\Downarrow\,\rangle =
a(\lambda)\,|\Downarrow\,\rangle,
}
\end{eqnarray}
where the functions $a(\lambda)$ and  $d(\lambda)$ are
equal to
\begin{equation} \label{ad}
a(\lambda) = \prod_{k=1}^{N} \sinh(\lambda-\nu_k+\eta), \qquad
d(\lambda) = \prod_{k=1}^{N} \sinh(\lambda-\nu_k-\eta).
\end{equation}
The vectors $\langle\,\Uparrow|$ and $\langle\,\Downarrow|$,
dual to (\ref{ud}), are eigenvectors of the operators
$A(\lambda)$ and $D(\lambda)$ with the same eigenvalues as in
equations (\ref{updown}) while instead of equations (\ref{die})
one has
\begin{equation} 
\langle\,\Uparrow|\,B(\lambda) = 0,\qquad
\langle\,\Downarrow|\,C(\lambda) = 0.
\end{equation}

Consider vectors generated by multiple action of operators
$ B(\lambda_\alpha) $ on the state $|\!\Uparrow\,\rangle$
\begin{equation} \label{BBB}
B(\lambda_M)\dots B(\lambda_1)\,|\Uparrow\,\rangle ,
\qquad  M \le N .
\end{equation}
The result of the action of the operator $A(\lambda)$ on vector (\ref{BBB})
follows from commutation relations (\ref{commAB})
\begin{equation} \label{ABBB}
\fl
A(\lambda)\,
\prod_{\alpha=1}^{M} B(\lambda_\alpha)\,|\Uparrow\,\rangle
=\Lambda
\prod_{\alpha=1}^{M} B(\lambda_\alpha)\,|\Uparrow\,\rangle
+\sum_{\beta=1}^{M} \Lambda_\beta
B(\lambda)
\prod_{\alpha=1\atop \alpha\ne\beta}^{M}
B(\lambda_\alpha) \,|\Uparrow\,\rangle ,
\end{equation}
where the coefficients $ \Lambda $ and $ \Lambda_\beta $ are
\begin{equation} 
\fl
\Lambda
=a(\lambda) \prod_{\gamma=1}^{M} f(\lambda,\lambda_\gamma),\qquad
\Lambda_\beta
=a(\lambda_\beta)\, g(\lambda_\beta,\lambda)
\prod_{\gamma=1\atop \gamma\ne\beta}^{M}
f(\lambda_\beta,\lambda_\gamma).
\end{equation}

The partition function
$ Z_N = Z_N (\lambda_1, \ldots ,\lambda_N; \nu_1, \ldots ,\nu_N) $
of the model on an $ N \times N $ square lattice with DWBC is obtained
by summation over the contributions of all possible spin configurations.
The contribution of each configuration is equal to the product of all vertex
weights of this configuration. In terms of QISM the partition function
may be represented as
\begin{equation} 
\fl
Z_N
=\left( \ot\limits_{\alpha=1}^{N}
{_\alpha}\langle \,\uparrow | \right) \otimes
\left( \ot\limits_{k=1}^{N}
{_k}\langle \,\downarrow | \right)
T_{N}(\lambda_N) \ldots T_{1}(\lambda_1)
\left( \ot\limits_{k=1}^{N}
|\uparrow\,\rangle_{k} \right) \otimes
\left( \ot\limits_{\alpha=1}^{N}
|\downarrow\,\rangle_{\alpha} \right) .
\end{equation}
The boundary conditions on the left and the right boundaries
extract from each matrix $ {\sf T}_\alpha(\lambda_\alpha) $
the operator $B(\lambda_{\alpha})$.
The boundary conditions on the top (bottom) of the lattice
correspond to the vector $|\Uparrow\,\rangle$ ($\langle\,\Downarrow|$),
respectively. Hence, the partition
function can be written in the form
\begin{equation} \label{Z}
Z_N = \langle\,\Downarrow|\,
B(\lambda_N) \dots  B(\lambda_1)\, |\Uparrow\,\rangle .
\end{equation}
Due to relation (\ref{commAB}) the order of operators $B(\lambda_\alpha)$
in the product is not essential.

The determinant representation for the partition function $Z_N$
was obtained in the papers \cite{I-87,ICK-92} and has the form
\begin{equation} \label{Z=detZ}
Z_N
=\frac{\prod\limits_{\alpha=1}^N \prod\limits_{k=1}^N
\sinh(\lambda_\alpha-\nu_k+\eta)\,\sinh(\lambda_\alpha-\nu_k-\eta)}
{\prod\limits_{1\leq \alpha<\beta \leq N}
\sinh(\lambda_\beta-\lambda_\alpha)
\prod\limits_{1\leq k < j\leq N} \sinh(\nu_k - \nu_j)} \,
{\det}_N {\cal Z}.
\end{equation}
The entries of the matrix ${\cal Z}$ are given by
\begin{equation} \label{matrixZ}
{\cal Z}_{\alpha k}=\phi(\lambda_\alpha,\nu_k), \qquad
\alpha,k =1,\dots,N
\end{equation}
where the function $\phi(\lambda,\nu)$ is
\begin{equation} \label{phi}
\phi(\lambda,\nu) =
\frac{\sinh 2\eta}{\sinh(\lambda-\nu+\eta)\sinh(\lambda-\nu-\eta)}.
\end{equation}
The proof of determinant representation (\ref{Z=detZ}) based
exclusively on commutation relations (\ref{commAB})
is given in Section 4.

\section{Boundary correlation functions}

In the present paper we consider two kinds of
correlation functions describing the local state probabilities at the
boundary. The first correlation function describes the probability of
absence of vertical solid line between $M+1$-th and $M$-th rows on
the first column and is known as ``boundary spontaneous polarization''.
In terms of QISM it is the one-point correlation
function of the local spin projector $q_1=\case 12 (1-\sigma_1^z)$ on
the spin down state, and it can be written as
\begin{equation} \label{corr1}
\fl
G_{N}^{(M)}
=Z_N^{-1}\,\langle\,\Downarrow|\, B(\lambda_N)\dots B(\lambda_{M+1})
\,q_1\, B(\lambda_M) \dots B(\lambda_1)
\,|\Uparrow\,\rangle.
\end{equation}
The second correlation function describes the probability 
that the solid line on the first column
turns to the left just on $M$-th row
\begin{equation} \label{corr2}
\fl
H_{N}^{(M)}
= Z_N^{-1}\,\langle\,\Downarrow|\,
B(\lambda_N)\dots B(\lambda_{M+1})\,q_1\,
B(\lambda_M)\,p_1\, B(\lambda_{M-1})\dots B(\lambda_1)
\,|\Uparrow\,\rangle,
\end{equation}
where $p_1$ is the projector on
the spin up state, $p_1=\case12 (1+\sigma_1^z)$.
Since $p_1+q_1=I$, these correlation
functions are related to each other as follows
\begin{eqnarray} \label{connection}
G_{N}^{(M)}=  H_{N}^{(M)}+H_{N}^{(M-1)}+\dots +H_{N}^{(1)},
\\ \label{connection2}
H_{N}^{(M)}= G_{N}^{(M)} - G_{N}^{(M-1)}.
\end{eqnarray}
However, it is easier to calculate them from definitions (\ref{corr1})
and (\ref{corr2}) independently.

In this Section we express the correlation functions of the model
on an $ N \times N $ square lattice through the sum over partition functions
of the models on $(N-1)\times(N-1)$ square sublattices.
We will call the corresponding formulae the ``reduction formulae''.
The derivation of these formulae is based exclusively on commutation
relations (\ref{commAB}). Since $ G_N^{(N)} =1 $, in the particular case
$ M=N $ the reduction formula for $ G_N^{(M)} $ turns into the recursion
relation for the partition function.
In the next Section we will prove that determinant
representation (\ref{Z=detZ}) is the solution of this recursion
relation. The determinant representation for the correlation functions can be
obtained then by substituting expression (\ref{Z=detZ})
in the reduction formulae, what makes our algebraic approach self-containing.

To derive the reduction formulae for the correlation functions we rewrite them
in the form suitable for applying commutation relations (\ref{commAB}).
Let us decompose the monodromy matrix
${\sf T}_\alpha(\lambda_\alpha)$
into the matrix product of two monodromy matrices in $\alpha$-th space
\begin{equation} \label{TT}
{\sf T}_\alpha(\lambda_\alpha) =
{\sf T}_{\alpha 2}(\lambda_\alpha)
{\sf T}_{\alpha 1}(\lambda_\alpha).
\end{equation}
This decomposition of the monodromy matrix is known
as the ``two-site model'' \cite{KBI-93}. In our case we choose these
matrices defined as follows
\begin{equation} \label{T2T1}
\eqalign{
{\sf T}_{\alpha 2}(\lambda_\alpha) =
{\sf L}_{\alpha N} (\lambda_\alpha,\nu_N)\dots
{\sf L}_{\alpha 2} (\lambda_\alpha,\nu_2) =
\pmatrix{A_2(\lambda_\alpha) & B_2(\lambda_\alpha) \cr
C_2(\lambda_\alpha) & D_2(\lambda_\alpha)}_{[\alpha]} ,
\\
{\sf T}_{\alpha 1}(\lambda_\alpha) =
{\sf L}_{\alpha 1} (\lambda_\alpha,\nu_1) =
\pmatrix{A_1(\lambda_\alpha) & B_1(\lambda_\alpha) \cr
C_1(\lambda_\alpha) & D_1(\lambda_\alpha)}_{[\alpha]}.
}
\end{equation}
The matrix elements of ${\sf T}_{\alpha 2}(\lambda)$
commute with the matrix elements of ${\sf T}_{\alpha 1}(\lambda)$ since
they are operators that act nontrivially in different spaces.
The entries of ${\sf T}_{\alpha 1}(\lambda)$ act nontrivially in the
first ``vertical'' space and depend on the ``vertical'' variable $\nu_1$,
while the entries of ${\sf T}_{\alpha 2}(\lambda)$
act nontrivially in the rest $N-1$ ``vertical'' spaces and
depend on the ``vertical'' variables $\nu_2,\dots,\nu_N$.
Each set of operators entering the monodromy matrices satisfy the
commutation relations given by (\ref{RTT}) and, in particular,
(\ref{commAB}). It should be noted that the generating vectors
$\,|\Uparrow\,\rangle$ ($\,|\Downarrow\,\rangle$) can be represented
as the direct product of two generating vectors, e.g.,
$\,|\Uparrow\,\rangle =\,|\Uparrow_2\,\rangle\otimes|\Uparrow_1\,\rangle$,
where $|\Uparrow_2\,\rangle=\otimes_{k=2}^{N}|\uparrow\,\rangle_k$
and $|\Uparrow_1\,\rangle\equiv|\uparrow\,\rangle_1$.
The properties of the both states
$\,|\Uparrow_2\,\rangle$ ($\,|\Downarrow_2\,\rangle$)
and $|\Uparrow_1\,\rangle$ ($\,|\Downarrow_1\,\rangle$)
are the same as of $|\Uparrow\,\rangle$ ($\,|\Downarrow\,\rangle$),
see the previous Section. The eigenvalue of, e.g., operator
$A_2(\lambda)$ on the state $|\Uparrow_2\,\rangle$ is
\begin{equation} \label{A2}
A_2(\lambda)\,|\Uparrow_2\,\rangle
=a_2(\lambda)\,|\Uparrow_2\,\rangle,\qquad
a_2(\lambda)
=\prod_{k=2}^{N} \sinh(\lambda-\nu_k+\eta).
\end{equation}
The decomposion of the operators
$A(\lambda)$, $B(\lambda)$, $C(\lambda)$ and $D(\lambda)$
follows from definitions (\ref{TT}), (\ref{T2T1}) and (\ref{T}).
In particular, one has
\begin{equation} \label{Btwosite}
B(\lambda) = A_2(\lambda)B_1(\lambda) +B_2(\lambda)D_1(\lambda).
\end{equation}
The operators $B_1(\lambda)$ and $D_1(\lambda)$ are the corresponding
entries of ${\sf L}$-operator (\ref{L}) and they are given by
\begin{equation} \label{B1D1}
B_1(\lambda) = \sigma_1^- \sinh 2\eta,\qquad
D_1(\lambda)=\sinh(\lambda-\nu_1-\eta\sigma_1^z).
\end{equation}
Using formulae (\ref{Btwosite}) and (\ref{B1D1}) one can reduce the problem
of calculation of the scalar products in the right hand sides of
expressions (\ref{corr1}) and (\ref{corr2}) to the problem of
calculation of the scalar products involving the operators
$A_2(\lambda)$ and $B_2(\lambda)$ only. Now we are ready to derive
the reduction formulae for the correlation functions.

We start with the derivation of the reduction formula
for the function $H_N^{(M)}$ since this derivation is straightforward.
Let us substitute expression (\ref{Btwosite}) in (\ref{corr2}) and
calculate the scalar product with respect to $\langle\,\Downarrow_1|$
and $|\Uparrow_1\,\rangle$ using formula (\ref{B1D1}).
We are left with the expression
\begin{eqnarray} \label{Hstep1}
\fl
H_{N}^{(M)}
=Z_N^{-1}\,\sinh2\eta
\prod_{\alpha=1}^{M-1}
\sinh(\lambda_{\alpha} - \nu_1 -\eta)
\prod_{\alpha=M+1}^{N}
\sinh(\lambda_{\alpha} - \nu_1 +\eta)
\nonumber\\
\lo
\times\langle\,\Downarrow_2|\,
B_2(\lambda_N) \dots B_2(\lambda_{M+1})\, A_2(\lambda_M)\,
B_2(\lambda_{M-1}) \dots B_2(\lambda_1)\,
|\Uparrow_2\,\rangle .
\end{eqnarray}
Applying (\ref{ABBB}) and taking into account (\ref{A2}),
we reduce the scalar product in (\ref{Hstep1})
to the sum over scalar products that involve only the operators $B_2$.
Comparing these scalar products with expression (\ref{Z}),
one immediately gets the following formula for the correlation function
$H_{N}^{(M)}$:
\begin{eqnarray} \label{Hstep2}
\fl
H_{N}^{(M)}
=Z_N^{-1}\,\sinh2\eta
\prod_{\alpha=1}^{M-1}
\sinh(\lambda_{\alpha} - \nu_1 -\eta)
\prod_{\alpha=M+1}^{N}
\sinh(\lambda_{\alpha} - \nu_1 +\eta)
\nonumber\\
\lo
\times\sum_{\beta =1}^{M}
a_2(\lambda_\beta)\,\frac{g(\lambda_\beta,\lambda_M)}
{f(\lambda_\beta,\lambda_M)}
\prod_{\gamma=1 \atop \gamma \ne \beta}^{M}
f(\lambda_\beta,\lambda_\gamma)\,
Z_{N-1}\!\left(\{\lambda_\alpha\}_{\alpha=1,\alpha\ne\beta}^N;
\{\nu_k\}_{k=2}^N\right).
\end{eqnarray}
In derivation of this formula we have used the fact that
the functions $ f(\lambda', \lambda)$ and $ g(\lambda', \lambda)$
defined in (\ref{fg}) satisfy the condition
$g(\lambda,\lambda)/f(\lambda,\lambda)=1$.
This allows us to include the first term arising in
(\ref{ABBB}) into the sum over $\beta$.

Equation (\ref{Hstep2}) expresses the particular
boundary correlation function of the model on an $N\times N$ square lattice
as the sum over partition functions of the models on $(N-1)\times (N-1)$
square sublattices. It should be stressed that in derivation of the formula
we have used only the algebra of operators $A_2(\lambda)$
and $B_2(\lambda)$ given by commutation relations (\ref{commAB}).

Let us turn now to the correlation function $G_N^{(M)}$ defined
by expression (\ref{corr1}). At first, we substitute formulae
(\ref{Btwosite}) and (\ref{B1D1}) in (\ref{corr1}). This gives
\begin{eqnarray} \label{Gstep1}
\fl
G_{N}^{(M)}
=Z_N^{-1}
\sinh2\eta
\sum_{\beta=1}^{M}
\prod_{\alpha=1}^{\beta-1}
\sinh(\lambda_\alpha - \nu_1 -\eta)
\prod_{\alpha=\beta+1}^{N}
\sinh(\lambda_\alpha - \nu_1 +\eta)
\nonumber\\
\lo\times
\langle\,\Downarrow_2|\,
B_2(\lambda_N) \dots B_2(\lambda_{\beta+1})\, A_2(\lambda_\beta)\,
B_2(\lambda_{\beta-1}) \dots B_2(\lambda_1)\,
|\Uparrow_2\,\rangle.
\end{eqnarray}
To obtain the representation of the correlation function $G_{N}^{(M)}$
as the sum over the partition functions of the models on
$(N-1)\times(N-1)$ square sublattices one can substitute
formula (\ref{ABBB}) in expression (\ref{Gstep1}), but then
the double sum will appear in the resulting expression.
To represent the right hand side of expression (\ref{Gstep1}) as a
single sum, like in formula (\ref{Hstep2}), we choose another way.
In expression (\ref{Gstep1}) we pick up the term which contains the
operator $A_2$ standing outmost to the left among the operators $B_2$,
i.e., the term corresponding to $\beta=M$.
Taking into account formula (\ref{ABBB}) it is
easy to see that this term contains the following contribution, which
corresponds to the first term in the right hand side of (\ref{ABBB}):
\begin{eqnarray} \label{cont}
\fl
\langle\,\Downarrow_2|\,
B_2(\lambda_N) \dots B_2(\lambda_{M+1})\, A_2(\lambda_M)\,
B_2(\lambda_{M-1}) \dots B_2(\lambda_1)\,
|\Uparrow_2\,\rangle
\nonumber\\
\fl
=a_2(\lambda_M) \prod_{\gamma=1}^{M-1} f(\lambda_M,\lambda_\gamma)\,
\langle\,\Downarrow_2|\,
B_2(\lambda_N) \dots B_2(\lambda_{M+1})\,
B_2(\lambda_{M-1}) \dots B_2(\lambda_1)\,
|\Uparrow_2\,\rangle
\nonumber\\
+\mbox{other terms},
\end{eqnarray}
where the ``other terms'' are the terms with scalar products
involving the operator $B_2(\lambda_M)$. Substituting equation (\ref{cont})
in (\ref{Gstep1}) and applying formula (\ref{ABBB}) to
the rest terms in (\ref{Gstep1}), we obtain that
\begin{eqnarray} \label{cont2}
\fl
G_{N}^{(M)}
=Z_N^{-1}
\prod_{\alpha=1}^{M}
\sinh(\lambda_\alpha - \nu_1 -\eta)\,
\prod_{\alpha=M+1}^{N}
\sinh(\lambda_\alpha - \nu_1 +\eta)
\nonumber\\
\fl\times
\Biggl(\frac{a_2(\lambda_M)\,\sinh2\eta}{\sinh(\lambda_M-\nu_1-\eta)}
\prod_{\gamma=1}^{M-1} f(\lambda_M,\lambda_\gamma)
\nonumber\\
\lo\times
\langle\,\Downarrow_2|\,
B_2(\lambda_N) \dots B_2(\lambda_{M+1})\,
B_2(\lambda_{M-1}) \dots B_2(\lambda_1)\,
|\Uparrow_2\,\rangle
+\mbox{other terms}
\Biggr),
\end{eqnarray}
where the ``other terms'' are again the terms, different
to those in (\ref{cont}), with scalar products involving the
operator $B_2(\lambda_M)$. Thus, the contribution written explicitly in
(\ref{cont2}) is the only possible one which does not contain
the operator $B_2(\lambda_M)$ in the scalar product.
Obviously, this contribution is symmetric under the permutations of
the elements in the set $ \lambda_1,\dots,\lambda_{M-1} $.
At the same time the correlation function $G_N^{(M)}$ is symmetric
under the permutations of the elements in the set
$\lambda_1,\dots,\lambda_M$, due to the commutativity of the operators
$B_2(\lambda_\alpha)$. It follows immediately from these symmetry
considerations that the whole expression standing in the parentheses in
(\ref{cont2}) is just the sum over the cyclic permutations of the elements
in the set $\lambda_1,\dots,\lambda_M$ of the term that written down
explicitly in the right hand side of equation (\ref{cont2}).
Therefore, the following representation
for the correlation function $ G_{N}^{(M)} $ is valid
\begin{eqnarray} \label{Gstep2}
\fl
G_{N}^{(M)}
=Z_N^{-1}\,
\prod_{\alpha=1}^{M}
\sinh(\lambda_{\alpha} - \nu_1 -\eta)
\prod_{\alpha=M+1}^{N}
\sinh(\lambda_{\alpha} - \nu_1 +\eta)
\nonumber\\
\lo
\times\sum_{\beta =1}^{M}
\frac{a_2(\lambda_\beta)\,\sinh2\eta}{\sinh(\lambda_\beta-\nu_1-\eta)}
\prod_{\gamma=1 \atop \gamma \ne \beta}^{M}
f(\lambda_\beta,\lambda_\gamma)\,
Z_{N-1}\!\left(\{\lambda_\alpha\}_{\alpha=1,\alpha\ne\beta}^N;
\{\nu_k\}_{k=2}^N\right).
\end{eqnarray}
The procedure described above is similar to the derivation of equation
(\ref{ABBB}) from commutation relations (\ref{commAB}), see, e.g.,
\cite{KBI-93}.

The important point to be discussed now is the meaning of representation
(\ref{Gstep2}) at $M=N$. Since the left hand side of equation
(\ref{Gstep2}) is equal to one at $M=N$, it
turns into the recursion relation on the partition function. In this
relation $Z_N$ is expressed through the sum over the partition
functions $Z_{N-1}$
\begin{eqnarray} \label{ZZ}
\fl
Z_N\!\left(\{\lambda_\alpha\}_{\alpha=1}^N;
\{\nu_k\}_{k=1}^N\right)=
\sinh2\eta\sum_{\beta =1}^{N}
\prod_{\alpha=1\atop\alpha\ne\beta}^{N}
\sinh(\lambda_{\alpha} - \nu_1 -\eta)
\prod_{k=2}^{N} \sinh(\lambda_\beta-\nu_k+\eta)
\nonumber\\
\lo
\times
\prod_{\gamma=1\atop\gamma\ne\beta}^{N}
f(\lambda_\beta,\lambda_\gamma)\times
Z_{N-1}\!\left(\{\lambda_\alpha\}_{\alpha=1,\alpha\ne\beta}^N;
\{\nu_k\}_{k=2}^N\right).
\end{eqnarray}
This formula was obtained by slightly different technique
in \cite{KMT-99}.
Due to the symmetry of $Z_N$ with respect to
the permutations of the variables $\{\nu_k\}_{k=1}^N$ one can rewrite this
recursion relation in a general form
\begin{eqnarray} \label{ZZj}
\fl
Z_N\!\left(\{\lambda_\alpha\}_{\alpha=1}^N;
\{\nu_k\}_{k=1}^N\right)=
\sinh2\eta\sum_{\beta =1}^{N}
\prod_{\alpha=1\atop\alpha\ne\beta}^{N}
\sinh(\lambda_{\alpha} - \nu_j -\eta)
\prod_{k=1\atop k\ne j}^{N} \sinh(\lambda_\beta-\nu_k+\eta)
\nonumber\\
\lo
\times
\prod_{\gamma=1\atop\gamma\ne\beta}^{N}
f(\lambda_\beta,\lambda_\gamma)\times
Z_{N-1}\!\left(\{\lambda_\alpha\}_{\alpha=1,\alpha\ne\beta}^N;
\{\nu_k\}_{k=1,k\ne j}^N\right),\qquad j=1,\dots,N.
\end{eqnarray}
Relations (\ref{ZZ}) and (\ref{ZZj}) are
valid for arbitrary values of the variables $\{\lambda_\alpha\}_{\alpha=1}^N$
and $\{\nu_k\}_{k=1}^N$.
Iterating relation (\ref{ZZj}) $ N-1 $ times, with the
initial condition $Z_1=\sinh2\eta$, one gets the answer
as the sum over all permutations $P$ of
$\{\lambda_\alpha\}_{\alpha=1}^N$ (c.f., (3.17) of \cite{B-87})
\begin{equation} \label{Z=perm}
\fl
Z_N=(\sinh2\eta)^N \sum_{P}^{}
\prod\limits_{1\leq\alpha<\beta\leq N}
f(\lambda_{P_\alpha},\lambda_{P_\beta})\,
\sinh(\lambda_{P_\beta}\!-\nu_{\alpha}-\eta)\,
\sinh(\lambda_{P_\alpha}\!-\nu_{\beta}+\eta).
\end{equation}
On the other hand, one can show that determinant representation
(\ref{Z=detZ}) for the partition function is the solution of
recursion relation (\ref{ZZj}). The determinant solution of equation
(\ref{ZZj}) leads to the determinant representations for the boundary
correlation functions.

\section{Determinant representations for boundary correlation functions}

We begin this Section with the proof that determinant formula
(\ref{Z=detZ}) for the partition function solves recursion relation
(\ref{ZZj}). The determinant representation for the boundary correlation
functions will follow then from reduction formulae (\ref{Hstep2}) and
(\ref{Gstep2}), which can be viewed as generalizations of the
recursion relation for the partition function.

Let us give the proof for $j=1$ in (\ref{ZZj}). This case is given by
relation (\ref{ZZ}). For the other values of $j$ the proof is
essentially the same. Consider the left hand side of relation
(\ref{ZZ}) which is given by formula (\ref{Z=detZ}). The aim is to
represent this expression in the form given by the right hand side of
relation (\ref{ZZ}). Since $Z_N$ and $Z_{N-1}$ are
expressed through determinants of $N\times N$ and $(N-1)\times (N-1)$
matrices respectively, it is clear that relation (\ref{ZZ})
can be proved developing a determinant of some $N\times N$ matrix
by the first column.
To apply this approach one has to transform the determinant of the matrix
$\cal Z$ in (\ref{Z=detZ}) first. Consider the function
\begin{equation} \label{gN}
g_N(\lambda)=
\frac{\prod\limits_{\gamma=1}^{N}
\sinh(\lambda_\gamma-\lambda+2\eta)}
{\prod\limits_{k=1}^{N}\sinh(\lambda-\nu_k-\eta)}.
\end{equation}
For this function one has the identity
\begin{equation} \label{g=sum}
g_N (\lambda_\alpha)=\sum_{k=1}^{N}\Phi_k\, \phi(\lambda_\alpha,\nu_k),
\qquad
\alpha=1,\dots,N,
\end{equation}
where the function $\phi(\lambda,\nu)$ is given
by formula (\ref{phi}). Here the coefficients $\Phi_k$
are independent of $\alpha$ and are given as
\begin{equation} 
\Phi_k = \frac{\prod\limits_{\gamma=1}^{N}
\sinh(\lambda_\gamma-\nu_k+\eta)}
{\prod\limits_{j=1\atop j\ne k}^{N}\sinh(\nu_k-\nu_j)},\qquad
k=1,\dots,N.
\end{equation}
Relation (\ref{g=sum}) is a short form of the identity
\begin{equation} \label{ind}
\fl
\frac{\prod\limits_{\gamma=1\atop \gamma\ne\alpha}^{N}
\sinh(\lambda_\gamma-\lambda_\alpha+2\eta)}
{\prod\limits_{j=1}^{N}\sinh(\lambda_\alpha-\nu_j-\eta)}
=\sum_{k=1}^{N}
\frac{\prod\limits_{\gamma=1\atop \gamma\ne\alpha}^{N}
\sinh(\lambda_\gamma-\nu_k+\eta)}
{\prod\limits_{j=1\atop j\ne k}^{N}\sinh(\nu_k-\nu_j)}\,
\frac{1}{\sinh (\lambda_\alpha-\nu_k-\eta)},
\end{equation}
which may be proved, e.g., by induction, see also \cite{G-83}.
Relation (\ref{g=sum}) may be considered as the system of linear
equations on the coefficients $\Phi_k$
with the right hand side formed by the vector $g_N(\lambda_\alpha)$,
$\alpha=1,\dots,N$. Therefore, by Cramer's rule we obtain the connection
formula for the determinants
\begin{equation} \label{detZ=det}
{\det}_N {\cal Z} = \frac{1}{\Phi_1}
\left|\matrix{
g_N(\lambda_1) & \phi(\lambda_1,\nu_2)& \dots &\phi(\lambda_1,\nu_N)\cr
g_N(\lambda_2) & \phi(\lambda_2,\nu_2)& \dots &\phi(\lambda_2,\nu_N)\cr
\vdots & \vdots & \ddots & \vdots\cr
g_N(\lambda_N) & \phi(\lambda_N,\nu_2)& \dots &\phi(\lambda_N,\nu_N)}
\right|_{N},
\end{equation}
where ${\det}_N {\cal Z}$ is given by formula (\ref{Z=detZ}).
It is the transformed form of the determinant of the matrix $\cal Z$
which can be developed by the first column. Hence, one gets
\begin{equation} \label{detZ=Delta}
{\det}_N {\cal Z} =
\frac{1}{\Phi_1} \sum_{\beta=1}^{N} (-1)^{\beta - 1} g_N(\lambda_\beta)
\Delta^{(\beta)}_{N-1}
\end{equation}
where $\Delta^{(\beta)}_{N-1}$ denotes the determinant
of an $(N-1)\times(N-1)$ matrix
\begin{equation} \label{Delta}
\Delta^{(\beta)}_{N-1} =
\left|\matrix{\phi(\lambda_1,\nu_2)& \dots &\phi(\lambda_1,\nu_N)\cr
\vdots & \ddots & \vdots\cr
\phi(\lambda_{\beta-1},\nu_2)& \dots &\phi(\lambda_{\beta-1},\nu_N)\cr
\phi(\lambda_{\beta+1},\nu_2)& \dots &\phi(\lambda_{\beta+1},\nu_N)\cr
\vdots & \ddots & \vdots\cr
\phi(\lambda_N,\nu_2)& \dots &\phi(\lambda_N,\nu_N)}
\right|_{N-1}.
\end{equation}
Substituting determinant representation (\ref{Z=detZ}) for the partition
functions $Z_N$ and $Z_{N-1}$ in the left and right hand sides
of relation (\ref{ZZ}) respectively, and cancelling the resulting
common factors, one gets exactly equation (\ref{detZ=Delta}).
This proves that the determinant representation (\ref{Z=detZ}) is the
solution of recursion relation (\ref{ZZ}) for the partition function.

Let us turn now to the boundary correlation functions.
Reduction formulae (\ref{Hstep2}) and (\ref{Gstep2}),
obtained in the previous Section, express them through the sum over
partition functions of the models on $(N-1)\times(N-1)$ square sublattices.
Using determinant representation (\ref{Z=detZ}) for the partition
function one can obtain then the determinant representations
for the boundary correlation functions.

Consider the correlation function $H_N^{(M)}$.
Substituting expression (\ref{Z=detZ}) for $Z_{N-1}$ in
reduction formula (\ref{Hstep2})
and extracting a general multiplier out of the sum over $\beta$ one
gets the following expression
\begin{eqnarray} \label{Hstep3}
\fl
H_{N}^{(M)}
=Z_N^{-1}\,\sinh2\eta
\prod_{\alpha=1}^{M-1}
\sinh(\lambda_{\alpha} - \nu_1 -\eta)
\prod_{\alpha=M+1}^{N}
\sinh(\lambda_{\alpha} - \nu_1 +\eta)
\nonumber\\
\lo
\times\frac{\prod\limits_{\alpha=1}^N \prod\limits_{k=2}^N
\sinh(\lambda_\alpha-\nu_k+\eta)\, \sinh(\lambda_\alpha-\nu_k-\eta)}
{\prod\limits_{1\leq \alpha<\beta \leq N}
\sinh(\lambda_\beta-\lambda_\alpha)
\prod\limits_{2\leq k < j\leq N} \sinh(\nu_k - \nu_j)} \,
\nonumber\\
\lo
\times\sum_{\beta=1}^{M}
\frac{\prod\limits_{\gamma=1}^{M-1}
\sinh(\lambda_\gamma-\lambda_\beta+2\eta)
\prod\limits_{\alpha=M+1}^{N}\sinh(\lambda_\gamma-\lambda_\beta)}
{\prod\limits_{k=2}^{N}\sinh(\lambda_\beta-\nu_k-\eta)}\,
(-1)^{\beta-1} \Delta^{(\beta)}_{N-1},
\end{eqnarray}
where the quantity $\Delta^{(\beta)}_{N-1}$ is defined by
expression (\ref{Delta}). Clearly, the sum over $\beta$ in formula
(\ref{Hstep3}) is nothing but the determinant of some $N\times N$ matrix
developed by the first column.
Only $M$ first entries in this column are not equal to zero so there
are $M$ terms in the sum. Taking this into account and using
expression (\ref{Z=detZ}) for $Z_N$ we finally obtain the following
determinant representation for the correlation function $H_N^{(M)}$:
\begin{equation} \label{H_N}
H_N^{(M)}
=\frac{\sinh2\eta\prod\limits_{k=2}^{N}\sinh(\nu_1-\nu_k)}
{\prod\limits_{\alpha=1}^{M}
\sinh(\lambda_{\alpha} - \nu_1 +\eta)
\prod\limits_{\alpha=M}^{N}
\sinh(\lambda_{\alpha} - \nu_1 -\eta)}
\, \frac{{\det}_N {\cal H}}{{\det}_N {\cal Z}},
\end{equation}
where the entries of the $N\times N$ matrix ${\cal H}$ are given by
\begin{equation} 
\eqalign{
{\cal H}_{\alpha 1} = h_M(\lambda_\alpha),\qquad
{\cal H}_{\alpha k} = \phi(\lambda_\alpha,\nu_k),\qquad k=2,\dots,N.
}
\end{equation}
The function $h_M(\lambda)$ is equal to
\begin{equation} \label{h_M}
h_M(\lambda)=
\frac{\prod\limits_{\gamma=1}^{M-1}
\sinh(\lambda_\gamma-\lambda+2\eta)
\prod\limits_{\gamma=M+1}^{N}\sinh(\lambda_\gamma-\lambda)}
{\prod\limits_{k=2}^{N}\sinh(\lambda-\nu_k-\eta)}.
\end{equation}
The points $\lambda_{M+1},\dots,\lambda_N$ are zeros of the
function $h_M(\lambda)$, hence, the last $N-M$ entries in the first
column of the matrix $\cal H$ are equal to zero. Note that the
prefactor in (\ref{H_N}) is, in fact,
equal to $\phi(\lambda_M,\nu_1)/h_M(\nu_1+\eta)$.

Consider now the correlation function $G_N^{(M)}$.
Substituting expression (\ref{Z=detZ}) for $Z_{N-1}$ in reduction formula
(\ref{Gstep2}) we arrive at the expression for the correlation
function $G_N^{(M)}$ as the sum of $\Delta_{N-1}^{(\beta)}$.
This expression is similar to (\ref{Hstep3})
for the function $H_N^{(M)}$. Finally, we obtain the following
determinant representation for the correlation function $G_N^{(M)}$:
\begin{equation} \label{G_N}
G_N^{(M)}
=\frac{\prod\limits_{k=2}^{N}\sinh(\nu_1-\nu_k)}
{\prod\limits_{\alpha=1}^{M}
\sinh(\lambda_{\alpha} - \nu_1 +\eta)
\prod\limits_{\alpha=M+1}^{N}
\sinh(\lambda_{\alpha} - \nu_1 -\eta)}
\, \frac{{\det}_N {\cal G}}{{\det}_N {\cal Z}},
\end{equation}
where the entries of the $N\times N$ matrix ${\cal G}$ are given by
\begin{equation} 
\eqalign{
{\cal G}_{\alpha 1} = g_M(\lambda_\alpha),\qquad
{\cal G}_{\alpha k} = \phi(\lambda_\alpha,\nu_k),\qquad
k=2,\dots,N.
}
\end{equation}
The function $g_M(\lambda)$ is equal to
\begin{equation} \label{gM}
g_M(\lambda)=
\frac{\prod\limits_{\gamma=1}^{M}
\sinh(\lambda_\gamma-\lambda+2\eta)
\prod\limits_{\gamma=M+1}^{N}\sinh(\lambda_\gamma-\lambda)}
{\prod\limits_{k=1}^{N}\sinh(\lambda-\nu_k-\eta)}.
\end{equation}
At $M=N$ this function is exactly the function $g_N(\lambda)$
defined by formula (\ref{gN}).
The points $\lambda_{M+1},\dots,\lambda_N$ are zeros of the
function $g_M(\lambda)$, hence, the last $N-M$ entries in the first
column of the matrix $\cal G$ are equal to zero.  By the direct check it is
easy to get convinced that representations (\ref{H_N}) and
(\ref{G_N}) satisfy relations (\ref{connection}) and
(\ref{connection2}).

Determinant representations (\ref{H_N}) and (\ref{G_N})
for the boundary correlation functions $H_N^{(M)}$ and $G_N^{(M)}$
are the main results of the present paper.

\section{The free fermion case}

The ``free fermion'' condition implies the following restriction
on the vertex weights:
\begin{equation} \label{ff}
{\sf a}^2(\lambda_{\alpha},\nu_k) +
{\sf b}^2(\lambda_{\alpha},\nu_k) =
{\sf c}^2(\lambda_{\alpha},\nu_k), \qquad \alpha,k = 1, \ldots, N.
\end{equation}
This equality is satisfied if we put $\eta=\rmi\case{\pi}{4}$ in (\ref{abc}).
It is convenient to change the variables $\lambda_\alpha\to\rmi\lambda_\alpha$,
$\nu_k\to\rmi\nu_k$, and after the rescaling
${\sf a}\to-\rmi{\sf a}$, ${\sf b}\to-\rmi{\sf b}$,
${\sf c}\to-\rmi{\sf c}$ one gets the following
parametrization of the vertex weights:
\begin{equation} 
\eqalign{
{\sf a}(\lambda_\alpha,\nu_k)
=\sin\left(\lambda_\alpha-\nu_k+\case{\pi}4\right),\\
{\sf b}(\lambda_\alpha,\nu_k)
=\sin\left(\lambda_\alpha-\nu_k-\case{\pi}4\right),\\
{\sf c}(\lambda_\alpha,\nu_k)=1.
}
\end{equation}
Since under the condition $\eta=\rmi\case{\pi}{4}$ the determinant
in equation (\ref{Z=detZ}) becomes the Cauchy determinant,
the partition function for the free fermion case can be evaluated
explicitly:
\begin{equation} \label{Zfree}
\fl
Z_N\!\left(\{\lambda_\alpha\}_{\alpha=1}^N;\{\nu_k\}_{k=1}^N\right)=
\prod\limits_{1\leq\alpha < \beta\leq N}\cos(\lambda_\alpha-\lambda_\beta)
\prod\limits_{1\leq k < j\leq N}\cos(\nu_k-\nu_j).
\end{equation}
For the correlation functions $H_N^{(M)}$ and $G_N^{(M)}$ of the model
in the free fermion case it is worth to use directly representations
(\ref{Hstep2}) and (\ref{Gstep2}). Substituting formula (\ref{Zfree})
in representation (\ref{Hstep2}), one gets for the correlation
function $H_N^{(M)}$
\begin{eqnarray} \label{Hstart}
\fl
H_{N}^{(M)} =
\frac{\prod\limits_{\alpha=1}^{M-1}
\sin(\lambda_{\alpha} - \nu_1 -\case{\pi}4)
\prod\limits_{\alpha=M+1}^{N}
\sin(\lambda_{\alpha} - \nu_1 +\case{\pi}4)}
{\prod\limits_{j=1}^N \cos(\nu_j - \nu_1)}
\nonumber\\
\lo
\times\sum_{\beta=1}^{M}
\frac{\prod\limits_{j=2}^N \sin(\lambda_\beta-\nu_j+\case{\pi}{4})}
{\prod\limits_{\alpha=M}^N
\cos(\lambda_\alpha-\lambda_\beta)}
\prod_{\alpha=1 \atop \alpha \ne \beta}^M
\frac{1}{\sin(\lambda_{\alpha} - \lambda_{\beta})}.
\end{eqnarray}
The similar formula can be easily
written down for the correlation function $G_N^{(M)}$.

Consider the homogeneous limit of the correlation functions in the free fermion
case. The homogeneous model is obtained by putting the variables in each set
$\{\lambda_\alpha\}_{\alpha=1}^N$ and $\{\nu_k\}_{k=1}^N$ to be equal:

\begin{equation} \label{hom}
\lambda_1=\dots=\lambda_N\equiv\lambda,\qquad
\nu_1=\dots=\nu_N\equiv\nu.
\end{equation}
Since now all the vertex weights depend only on the difference $\lambda-\nu$,
without loss of generality one can take $\nu=-\case\pi4$.
Thus, we have the homogeneous model with the following parametrization
of the vertex weights
\begin{equation} \label{cs}
{\sf a}(\lambda)=\cos\lambda,\qquad
{\sf b}(\lambda)=\sin\lambda,\qquad
{\sf c}(\lambda)=1.
\end{equation}
Note that the partition function $Z_N$ of the homogeneous model is
independent of $\lambda$ and is equal to one, $Z_N=1$, see (\ref{Zfree}).

The substitution $\nu_k=\nu=-\case\pi4$ in expression
(\ref{Hstart}) leads us to the following
intermediate expression for $H_N^{(M)}$, when all $\nu_k$ are equal while all
$\lambda_\alpha$ are still different
\begin{equation} \label{Hla}
\fl
H_{N}^{(M)}=
\prod_{\alpha=1}^{M-1}\sin\lambda_\alpha
\prod_{\alpha=M+1}^{N}\cos\lambda_\alpha
\times
\sum_{\beta=1}^{M}
\frac{\cos^{N-1}\lambda_\beta}
{\prod\limits_{\alpha=M}^N \cos(\lambda_\alpha-\lambda_\beta) }
\prod\limits_{\alpha=1\atop\alpha\ne\beta}^{M}
\frac{1}{\sin(\lambda_\alpha-\lambda_\beta)}.
\end{equation}
The problem now is to obtain the limit of expression (\ref{Hla}) when
all $\lambda_\alpha$ tend to the same value.
For this purpose it is convenient to rewrite (\ref{Hla})
in terms of rational functions instead of trigonometric ones
\begin{equation} \label{Hu}
H_{N}^{(M)}=
(1+u_M^2)
\prod_{\alpha=1}^{M-1} u_\alpha\times
\sum_{\beta=1}^{M}
\prod\limits_{\alpha=M}^N \frac{1}{1+u_\alpha u_\beta}
\prod\limits_{\alpha=1\atop\alpha\ne\beta}^{M}
\frac{1}{u_\alpha - u_\beta} \, ,
\end{equation}
where
\begin{equation} 
u_\alpha=\tan\lambda_\alpha,\qquad \alpha=1,\dots,N.
\end{equation}
Therefore, one should take the homogeneous limit in the set
$\{u_{\alpha}\}_{\alpha=1}^{N}$:
\begin{equation} 
u_{\alpha} \to u,  \qquad \alpha=1, \ldots, N .
\end{equation}
For the subset $\{u_{\alpha}\}_{\alpha=M}^{N}$ we may simply put
$u_M=u_{M+1}=\ldots=u_N=u$, while for the subset
$\{u_{\alpha}\}_{\alpha=1}^{M-1}$ we parametrize $u_\alpha$ as
\begin{equation} \label{ua}
u_\alpha=u-(M-\alpha)\varepsilon, \qquad \alpha=1, \ldots, M-1,
\end{equation}
and the homogeneous limit corresponds then to the case of vanishing
$ \varepsilon $. Since the prefactor of the sum over $ \beta $ in (\ref{Hu})
is regular as $ \varepsilon \to 0 $, one may put $ \varepsilon = 0 $ in
this perfactor.
Thus, for $ H_{N}^{(M)} $ one gets
\begin{eqnarray} 
\fl
H_{N}^{(M)}=
(1+u^2)
u^{M-1}
\sum_{\beta=1}^{M}
\frac{1}{[1+u^2-u(M-\beta)\varepsilon]^{N-M+1}\,\varepsilon^{M-1}}
\prod\limits_{\alpha=1\atop\alpha\ne\beta}^{M}
\frac{1}{\alpha -\beta}
\nonumber\\
\lo
=(1+u^2) u^{M-1}
\sum_{\beta=0}^{M-1}
\frac{(-1)^{M-1}}{(1+u^2-u\beta\varepsilon)^{N-M+1}\,\varepsilon^{M-1}}
\prod\limits_{\alpha=0\atop\alpha\ne\beta}^{M-1}
\frac{1}{\alpha -\beta}.
\end{eqnarray}
Taking into account that
\begin{equation} 
\prod\limits_{\alpha=0\atop\alpha\ne\beta}^{M-1}
\frac{1}{\alpha -\beta} = \frac{(-1)^\beta}{\beta!\,(M-\beta-1)!}
=\frac{(-1)^\beta}{(M-1)!} {M-1\choose \beta}
\end{equation}
one obtains
\begin{equation} \label{almostdone}
\fl
H_{N}^{(M)}=
\frac{(-1)^{M-1}(1+u^2)}{(M-1)!\,u^{N-2M+2}}
\sum_{\beta=0}^{M-1}
\frac{(-1)^\beta}{[(1+u^2)u^{-1}-\varepsilon\beta]^{N-M+1}\,
\varepsilon^{M-1}}
{M-1\choose \beta}.
\end{equation}

The sum over $ \beta $ in (\ref{almostdone}) in the limit
$ \varepsilon \to 0 $ becomes exactly the $ (M-1)$-th
derivative of a function $ f(z) $ with respect to a variable $ z $
\begin{equation} 
\lim_{\varepsilon\to 0}
\sum_{\beta=0}^{M-1}
\frac{(-1)^\beta f(z-\beta\varepsilon)}{\varepsilon^{M-1}}
{M-1\choose \beta}=
\frac{\rmd^{M-1}}{\rmd z^{M-1}} f(z),
\end{equation}
where
\begin{equation} 
f(z) =\frac{1}{z^{N-M+1}},\qquad z=\frac{1+u^2}{u}.
\end{equation}
Hence, for the sum over $ \beta $
in (\ref{almostdone}) in the limit $ \varepsilon \to 0 $ one gets
\begin{equation} \label{lim}
\fl
\lim_{\varepsilon\to 0}
\sum_{\beta=0}^{M-1}
\frac{(-1)^\beta}{[(1+u^2)u^{-1}-\varepsilon\beta]^{N-M+1}\,
\varepsilon^{M-1}} {M-1\choose \beta}
= \frac{(-1)^{M-1}(N-1)!}{(N-M)!}
\left(\frac{u}{1+u^2}\right)^N.
\end{equation}

Finally, substituting expression (\ref{lim}) in (\ref{almostdone}) and
taking into account that $u=\tan\lambda$ we obtain the following
expression for the correlation function $H_N^{(M)}$ of the homogeneous
six-vertex model in the free fermion case (\ref{cs}):
\begin{equation} \label{Hff}
H_{N}^{(M)}
={N-1\choose M-1}\,
\left(\cos^2\lambda\right)^{N-M} \left(\sin^2\lambda\right)^{M-1}.
\end{equation}

Similarly, for the correlation function $G_N^{(M)}$ one gets
\begin{equation} \label{Gff}
G_N^{(M)}= \sum_{K=1}^{M}
{N-1\choose K-1}\,
\left(\cos^2\lambda\right)^{N-K} \left(\sin^2\lambda\right)^{K-1}.
\end{equation}
Obviously, connection formulae (\ref{connection}) and (\ref{connection2})
are fulfilled by (\ref{Hff}) and (\ref{Gff}).

It is clear from the obtained results that the boundary correlation functions
significantly depend on both $\lambda$ and $M$ even in the simplest
free fermion case.
This dependence of the function $G_N^{(M)}$ is of especial interest 
in the thermodynamic limit when both $N$ and $M$ go to infinity so that
the variable $x=M/N$ runs through the interval $ 0< x <1 $.
Let us denote the function $G_N^{(M)}$ in this limit as $G(x)$.
Representing (\ref{Gff}) in the form
\begin{equation} \label{Gffint}
\fl
G_N^{(M)}=
{N-1 \choose M-1}\,(\sin^2 \lambda)^{M-1} (\cos^2 \lambda)^{N-M}\,
{}_{2}F_{1} (1,-M+1;N-M+1;-\cot^2 \lambda) 
\end{equation}
and using the integral representation for the hypergeometric function
${}_2F_1$, one gets
\begin{eqnarray} \label{Gintegral}
\fl G_N^{(M)} =
\frac{(N-1)!}{(N-M-1)!(M-1)!}
\left(\sin^2 \lambda\right)^{M-1}
\left(\cos^2 \lambda\right)^{N-M}
\nonumber\\
\lo
\times
\int\limits_{0}^{1}dt\, (1-t)^{N-M-1}
\left(1+t\cot^2 \lambda \right)^{M-1},\qquad M=1, \ldots, N-1.
\end{eqnarray}
Applying the steepest descent method to representation (\ref{Gintegral})
we obtain that $G(x)$ is the Heaviside step function:
\begin{equation} \label{step}
G(x) =\theta\left(x-\sin^2\lambda\right),\qquad
\theta(\xi)=
\cases{
1 &if $\xi>0$\\
\case{1}{2} & if $\xi=0$\\
0 & if $\xi<0$}.
\end{equation}
This result means that
the arrows at the boundary column are ordered (frozen) in the
thermodynamic limit: all arrows are pointing down above the point
$ x= \sin^2 \lambda $, while below all of them are pointing up.

\section{Determinant representations in the homogeneous limit}

In this Section the homogeneous limit of determinant
representations (\ref{H_N}) and (\ref{G_N}) for the 
boundary correlation functions is considered in the general case.
To obtain the homogeneous model one should put all
$\lambda_\alpha$ equal to $\lambda$ and all $\nu_k$ equal to $\nu$ in
the inhomogeneous model (\ref{abc}).
Without loss of generatity one may put also $\nu=0$.
Hence, the vertex weights of the homogeneous six-vertex model are given by
\begin{equation} 
{\sf a}(\lambda)=\sinh(\lambda +\eta), \qquad
{\sf b}(\lambda)=\sinh(\lambda -\eta), \qquad
{\sf c}(\lambda)=\sinh 2\eta.
\end{equation}
The procedure of taking the homogeneous limit
for the partition function $Z_N$ have been elaborated in
the papers \cite{I-87,ICK-92}. In this limit,
the singularities coming from the denominator
of expression (\ref{Z=detZ}) are cancelled by the zeroes coming
from the determinant since then all rows and
columns of the matrix ${\cal Z}$ tend to each other.

It can be shown by Taylor expansion of the entries of the matrix ${\cal Z}$
that the partition function of the homogeneous model is expressed through
the double Wronskian
\begin{equation}  \label{Zhom}
Z_{N} = \frac{\det_N{\cal Z}_{\rm hom}}
{\bigl[\phi(\lambda)\bigr]^{N^2}\prod\limits_{n=1}^{N-1} (n!)^2},\qquad
({\cal Z}_{\rm hom})_{\alpha k} =
\frac{\rmd^{\alpha+k-2}}{\rmd \lambda^{\alpha+k-2}}
\phi(\lambda),
\end{equation}
where $\phi(\lambda)\equiv \phi(\lambda,0)$, namely,
\begin{equation}  
\phi(\lambda)  =
\frac{\sinh 2\eta}{\sinh(\lambda+\eta)\sinh(\lambda-\eta)}.
\end{equation}
Since the correlation functions have been expressed in equations 
(\ref{H_N}) and (\ref{G_N}) through determinants, one may apply
the approach described in detail in \cite{ICK-92} to obtain the
correlation functions of the homogeneous model.

To apply the procedure given in \cite{ICK-92} with minimal
modifications, it is convenient to consider the
function ${\tilde G}_{N}^{(M)}$ instead of $G_N^{(M)}$, where
\begin{equation} \label{rotate}
{\tilde G}_{N}^{(M)}(\lambda_1,\dots,\lambda_N;\nu_1,\dots,\nu_N)
=G_{N}^{(M)}(\lambda_N,\dots,\lambda_1;\nu_N,\dots,\nu_1).
\end{equation}
Clearly, these functions are equal in the homogeneous limit.
Comparing equations (\ref{G_N}) and (\ref{rotate})
one can see that the function ${\tilde G}_{N}^{(M)}$
may be written as
\begin{equation}  \label{tildeG}
\lo
{\tilde G}_{N}^{(M)}=\frac{ \prod\limits_{k=1}^{N-1} \sinh(\nu_N-\nu_k)}
{\prod\limits_{\alpha=N-M+1}^{N} \sinh(\lambda_{\alpha} - \nu_N + \eta)
\prod\limits_{\alpha=1}^{N-M} \sinh(\lambda_{\alpha} - \nu_N - \eta)}
\frac{\det_N \tilde{\cal G}}{\det_N{\cal Z}} ,
\end{equation}
where
\begin{equation} 
\lo
{\tilde{\cal G}}_{\alpha k} = \phi(\lambda_\alpha, \nu_k) , \qquad
{\tilde{\cal G}}_{\alpha N} = \tilde g_{M}(\lambda_\alpha) , \qquad
k=1, \ldots, N-1 ,
\end{equation}
and the function ${\tilde g}_{M}(\lambda)$ is defined as
\begin{equation} 
\lo
{\tilde g}_{M}(\lambda) = \frac{\prod\limits_{\gamma=N-M+1}^N
\sinh(\lambda_{\gamma} - \lambda +2\eta)
\prod\limits_{\gamma=1}^{N-M} \sinh(\lambda_\gamma - \lambda)}
{\prod\limits_{k=1}^N \sinh(\lambda - \nu_k - \eta)}.
\label{GaN}
\end{equation}
The matrix $\tilde{\cal G}$ differs from the matrix ${\cal Z}$
by the last column, and the first $N-M$ entries in the last column
of the matrix $\tilde{\cal G}$ are equal to zero,
$\tilde{\cal G}_{N 1}=\ldots=\tilde{\cal G}_{N N-M}=0$.

Now, the homogeneous limit in the set
$\{\lambda_\alpha\}_{\alpha=1}^N$ can be easily found
following \cite{ICK-92}.
Representing the differences $\lambda_\gamma-\lambda_\alpha$
in the expression for the quantity $\tilde g_M(\lambda_\alpha)$
as $(\lambda_\gamma-\lambda)-(\lambda_\alpha-\lambda)$, one may
consider the differencies $\lambda_\alpha-\lambda$ as independent
variables. In the limit $\lambda_\alpha\to\lambda$,
$\alpha=1,\ldots,N$ these variables tend to zero
and it can be proved that the entries in the last column of the matrix
${\cal Z}$, after successive subtractions of the rows
become the coefficients in Taylor expansion of the function
\begin{equation} 
\psi(\varepsilon)=(-1)^N
\frac{(\sinh \varepsilon)^{N-M} \sinh^M (\varepsilon-2\eta)}
{\prod\limits_{j=1}^N\sinh(\varepsilon + \lambda -\nu_j - \eta)}
\end{equation}
at the point $\varepsilon = 0$. This solves the problem
of taking the homogeneous limit in the set
$\{\lambda_\alpha\}_{\alpha=1}^N$. The homogeneous limit in the set
$\{\nu_k\}_{k=1}^N $ can be taken in the same manner as for the
partition function, with the only difference that there are
no subtractions from the last column; one should simply put all
$\nu_k$ equal to $\nu=0$ in the entries of the last column.

As a result, one obtains the following
determinant representation for the correlation function
$G_N^{(M)}$ in the homogeneous limit:
\begin{equation} \label{Ghom}
G_N^{(M)} =
\frac{(N-1)!}
{[\sinh (\lambda +\eta)]^M
[\sinh(\lambda -\eta)]^{N-M}}
\frac{{\det}_N {\cal G}_\mathrm{hom}}{{\det}_N {\cal Z}_\mathrm{hom}} ,
\end{equation}
where
\begin{eqnarray}
\lo
({\cal G}_\mathrm{hom})_{\alpha k} = ({\cal Z}_\mathrm{hom})_{\alpha k},
\qquad k=1, \ldots, N-1,
\nonumber\\
\lo
({\cal G}_\mathrm{hom})_{\alpha N} =
- \frac{\rmd^{\alpha-1}}{\rmd \varepsilon^{\alpha -1}}
\left\{\frac{(\sinh \varepsilon)^{N-M} [\sinh(\varepsilon-2\eta)]^M}
{[\sinh(\varepsilon+\lambda-\eta)]^N}
\right\}\Bigg|_{\varepsilon =0} .
\end{eqnarray}
For the correlation function $H_N^{(M)}$
in the homogeneous limit the following expression is valid
\begin{equation}  \label{Hhom}
H_N^{(M)} =
\frac{(N-1)! \sinh 2\eta}
{[\sinh(\lambda+\eta)]^{M} [\sinh(\lambda-\eta)]^{N-M+1}}
\frac{\det_N {\cal H}_\mathrm{hom}}{\det_N {\cal Z}_\mathrm{hom}} ,
\end{equation}
where
\begin{eqnarray}
\lo
({\cal H}_\mathrm{hom})_{\alpha k} = ({\cal Z}_\mathrm{hom})_{\alpha k},
\qquad k=1, \ldots, N-1,
\nonumber\\
\lo
({\cal H}_\mathrm{hom})_{\alpha N} =
\frac{\rmd^{\alpha-1}}{\rmd \varepsilon^{\alpha -1}}
\left\{\frac{(\sinh \varepsilon)^{N-M}[\sinh(\varepsilon-2\eta)]^{M-1}}
{[\sinh(\varepsilon+\lambda-\eta)]^{N-1}}
\right\}\Bigg|_{\varepsilon =0} .
\end{eqnarray}
Formulae (\ref{Ghom}) and (\ref{Hhom}) generalize 
representation (\ref{Zhom}) for the partition function of the
homogeneous model, and they may be used for investigation of the
correlation functions in the thermodynamic limit.

\section{Conclusion}

In the present paper the determinant representation for the one-point
boundary correlation functions of the six-vertex model with the domain
wall boundary conditions is derived. For the correlation functions
$H_N^{{M}}$ (\ref{corr2}) and $G_N^{(M)}$ (\ref{corr1}) we have
obtained representations (\ref{H_N}) and (\ref{G_N}), which
generalize the determinant representation for the partition function
$Z_N$. In these formulae the correlation functions are expressed
through the determinants of the $N\times N$ matrices ${\cal H}$ and 
${\cal G}$ respectively. The matrices ${\cal H}$ and ${\cal G}$ 
differ from the matrix ${\cal Z}$ appearing in
representation (\ref{Z=detZ}) for the partition function by the first
column only. The derivation of representations (\ref{H_N}) and (\ref{G_N})
is based on reduction formulae (\ref{Hstep2}) and (\ref{Gstep2})
which have been obtained in Section 3 exclusively by means of the algebra
of operators entering the monodromy matrix.

In the free fermion case studied in Section 5 we have
obtained explicit (determinant-free) formulae (\ref{Hff}) and
(\ref{Gff}) for the boundary correlation functions
$H_N^{{M}}$ and $G_N^{(M)}$ of the homogeneous model.
In the thermodynamic limit $N,M\to\infty$
the function $G_N^{(M)}$ describing
the spontaneous polarization at the boundary turns into
the function $G(x)$ with $x= M/N$, $0<x<1$. We have found that
$G(x)$ is just the Heaviside step
function (\ref{step}). The emergence of the step function for the spontaneous
polarization indicates the ``freezing'' of the arrows at the boundary
in the thermodynamic limit.
On the other hand, from the results of the paper \cite{Z-96} we
have found numerically that at the ice point,
${\sf a}={\sf b}={\sf c}=1$, the function $G(x)$ exhibits the similar
behaviour, $G(x)=\theta(x-1/2)$.
The obtained determinant formulae (\ref{Ghom}) and (\ref{Hhom})
may be used for investigation of the boundary correlations
in the thermodynamic limit in the homogeneous model with arbitrary values
of the vertex weights.

The step function behaviour of the boundary spontaneous polarization
indicate the existence of the analogue of the arctic circle
theorem \cite{CEP-96,JPS-98}.
To be more precise, we expect that for the
values $-1<\cosh 2\eta<1$ the arrows are ``frozen'' at the corners of
the grid, while inside the grid there is a region of ``disorder''.
To obtain the shape of the ``disordered'' region of the grid it is
necessary to obtain an appropriate expression for the spontaneous
polarization not only at the boundary but also at the arbitrary point
of the lattice.
Hence, the problem of calculation of the correlation functions of
the model deserves further investigation. We hope that the approach
described in Sections 3 and 4 may appear to be productive
in the derivation of the correlation functions outside of the boundary.

\ack

We thank V. Tarasov for useful discussions.
This work was partially supported by the RFBR Grant No 01-01-01045.
One of us (A.G.P.) was partially supported by the INTAS Grant
No 99-01705 and by the CRDF Grant No RM1-2244.



\end{document}